\documentclass[prl,aps,showpacs,nofootinbib,floatfix,twocolumn ]{revtex4}
\usepackage{slashed}
\usepackage{float}
 \usepackage{amsmath,graphicx,color,epsfig,ulem}
 \usepackage{amssymb}

\begin{document}
\title{Finite volume effects with stationary wave solution from Nambu--Jona-Lasinio model}
\author{Qing-Wu Wang $^{1}$}~\email[]{Email: qw.wang@scu.edu.cn}
\author{Yonghui Xia$^{2}$}
\author{Hong-Shi Zong$^{2}$}~\email[]{Email: zonghs@nju.edu.cn}
\affiliation{
$^1$Department of Physics, Sichuan University,  Chengdu 610064, China\\
$^2$ Department of Physics, Nanjing University, Nanjing 210093, China\\
}

\begin{abstract}
In this paper, we use the two-flavor Nambu-Jona-Lasinio (NJL) model with the proper time regularization to study the finite-volume effects of QCD chiral phase transition.
 Within a cubic volume of finite size $L$,  we choose the stationary wave condition (SWC) as the real physical spatial boundary conditions of quark fields and compare our results with that by means of commonly used (anti-)period boundary condition (APBC or PBC). It is found that the results by means of SWC are obviously different to the results from the APBC or PBC.  Although the three boundary conditions give the same chiral crossover transition curve in the infinite volume limit, the limit size $L_0$ (when $L\geq L_{0}$, the chiral quark condensate $-\left\langle {  \bar \psi \psi} \right\rangle_L$ is indistinguishable from that at $L=\infty$) using SWC is  $L_0\approx 500$ fm which is much larger than the results obtained using APBC or PBC. More importantly,  $L_0\approx 500$ fm is also much large than the typical size of the quark-gluon plasma produced by the relativistic heavy ion collisions. This means that the finite volume effects play a very important role in Relativistic Heavy Ion Collisions. In addition, we also found that when $L\leq 2$ fm, even at zero temperature the chiral symmetry is effectively restored. Furthermore, to quantitatively reflect the finite volume effects on the QCD chiral phase transition, we introduce a new vacuum susceptibility, $\chi_{1/L}(T)=-\frac{\partial \left\langle {  \bar \psi \psi} \right\rangle}{\partial (1/L)}$. With this new vacuum susceptibility, it is very interesting to find $\chi_{1/L}(T=0)=\chi_{1/L}(T=1/L)$ for SWC.
\end{abstract}
\maketitle


Dynamical chiral symmetry breaking (DCSB) is one of the key feature of Quantum Chromodynamics (QCD).
The chiral phase transition at finite temperature is of continuous interests for studying the QCD phase diagram
\cite{Halasz:1998qr,Aoki:2006we,Bali:2012zg,Cui:2017ilj}. Many different methods have been used to analyse chiral symmetry breaking and restoration in variant situation.
 Other than color confinement, DCSB involving light degrees of freedoms which may  propagate over long distances is thus  closely relevant to the size of the system.
In the early universe, a few microseconds after big bang, when the temperature was extremely high, the quark-gluon plasma (QGP) may have been prevalent.
Experimentally, such a state can be reproduced  in laboratory  by relativistic heavy ion collisions (RHICs) \cite{Adams:2005dq,Shuryak:2008eq}.
The matter formed due to the energy deposition of the colliding heavy ion obviously has a finite volume.
Volume of homogeneity ranges between approximately $50 \sim 250$ fm$^3$ for Au-Au and Pb-Pb collisions space, while volume of the smallest QGP system produced is estimated to be
as low as (2 fm)$^3$ \cite{Bass:1998qm,Palhares:2009tf,Graef:2012sh}.
In view of the finite QGP size produced in RHICs  can be compared with the wavelength of $\pi$ meson, so it is very important to study phenomena related to the finite volume size.  Actually,  finite volume effects in QCD have already been studied for several  decades \cite{Klein:2017shl}. The steady improvements of lattice simulations also make the calculations on finite volume effects possible and to give accurate results a thorough understanding of finite volume effects is needed.

Many  different methods have been proposed to study the finite volume effects \cite{Fisher:1972zza,Gasser:1986vb,Gasser:1987zq,Braun:2004yk, Braun:2005fj,Colangelo:2005gd,Colangelo:2010ba, Luecker:2009bs,Li:2017zny}, and a recent summary is given in Ref.\cite{Klein:2017shl}. Within a finite volume, a concrete boundary condition needs to be chosen in advance. In the past, there are two typical boundary conditions: periodic boundary condition (PBC) and anti-periodic boundary condition (APBC), namely APBC for the quark fields and PBC for gluon fields.  At finite temperature, it is  claimed that  the particle field should take the same boundary condition (PBC or APBC) in the spatial and temporal directions to ensure permutation symmetry. The quark-meson model gives results consistent with chiral perturbation theory with APBC \cite{Klein:2017shl}.  Quark-meson model and lattice QCD simulation show results of low-energy behaviors depend on the choice of the quark boundary condition
\cite{Braun:2005fj,Carpenter:1984dd,Fukugita:1989yw,Aoki:1993gi}, even though the lattice simulation still takes the PBC  as a \textit{ de facto}
standard \cite{Aoki:1993gi,Klein:2017shl}.

Before we discuss the finite-volume effects on QCD chiral phase transition, a brief retrospect of  the finite volume effects on the black-body radiation is beneficial.
As we all know, when the size of the black body cavity is large enough, the black-body radiation spectrum does not depend on the choice of the spatial boundary condition.
That is, whether it is PBC, APBC or a stationary wave condition (SWC), none of the final results will be affected.
But when the size of the black-body cavity is small enough, to ensure that photon gas is confined to the cavity, people must use the SWC to study the finite volume effect on black body radiation. Therefore, in this article we will adopt the SWC to explore the finite volume effect on the QCD phase transition and compare our results with those by means of PBC and APBC used in the past.

We will calculate the finite-volume effects of QCD chiral phase transition with the three types of boundary in the framework of NJL model.
 The NJL model is a faithful phenomenological model of QCD \cite{Klevansky:1992qe,Buballa:2003qv}. It provides insight into  the quark flavor dynamics.   The Lagrangian is
\begin{eqnarray}
\mathcal{L}_{NJL} & =&\bar \psi (i\gamma_\mu)\partial^\mu-\hat m_q\psi  \nonumber\\
 &+&G[(\bar \psi  \psi)^2+(\bar \psi i\gamma_5\tau  \psi)^2],
\end{eqnarray}
where $G$ is the four-quark effective coupling. We consider only the u-d quark degree of freedom and work in the limit of exact isospin symmetry.

 In the mean field approximation, the effective quark mass is $  M=m+\sigma$ with
 \begin{equation}
 \sigma=-2G\left\langle {  \bar \psi \psi} \right\rangle
 \end{equation}
 and the chiral quark condensate is defined as
\begin{equation}\label{eq.gap1}
\left\langle {  \bar \psi \psi} \right\rangle=-\int \frac{d^4p}{(2\pi)^4}Tr[S(p)],
\end{equation}
where $S(p)$ is the dressed quark propagator and the trace is taken in color, flavor and Dirac spinor spaces.

Since the NJL model is non-renormalizable, a cut off on the  momentum  integration is usually implemented for regularization. There are many different regularization schemes and we will use the proper time regularization \cite{Cui:2017ilj,Braun:2004yk,Braun:2005fj,Liao:1994fp,Litim:2001hk,Zappala:2002nx,Cui:2014hya,Zhang:2016zto} here.
 Under this regularization scheme  the  trace term in Eq.(\ref{eq.gap1}) is replaced by an integral with
 a suitable choice of the cutoff function. Here in the gap equation  the key equation is a replacement $ \frac{1}{A(p^2)^n} \rightarrow \frac{1}{(n-1) !}\int_{\tau_{UV}}^ \infty d\tau \tau^{n-1}e^{-\tau A(p^2)}$. Then the chiral quark condensate in the infinite volume and  at zero temperature  can be written as
\begin{eqnarray}
\left\langle {  \bar \psi \psi} \right\rangle&=&-N_cN_f\int\frac{d^4p}{(2\pi)^4}\frac{4M}{p^2+M^2}  \nonumber \\
&=&-24M\int^\infty_{-\infty} \frac{d^4p}{(2\pi)^4}\int^ \infty_{\tau_{UV}} d\tau e^{-\tau(p^2+M^2)}  \nonumber\\
&=&-\frac{3M}{2\pi^2}\int^ \infty_{\tau_{UV}} d\tau\frac{ e^{-\tau M^2}}{\tau^2}.
\end{eqnarray}

At finite temperature, the quark four-momentum is  replaced by $p_k=(\vec{ p}, \omega_k)$, with $\omega_k=(2k+1)\pi T$, $k \in  \mathbb{Z}$ for fermion. The fourth momentum is replaced by a sum of all the fermion Matsubara frequencies $\omega_k$. Then the two quark condensate satisfies
\begin{eqnarray}
\left\langle {  \bar \psi \psi} \right\rangle&=&-24M\int^ \infty_{\tau_{UV}} d\tau e^{-\tau M^2}\times \nonumber\\
&&T\sum^\infty_{k=-\infty}\int^\infty_{0} \frac{dp}{2\pi^2}p^2e^{-\tau( p^2+\omega_k^2)}   \nonumber  \\
&=&-\frac{3MT}{\pi^{3/2}}\int^  \infty_{\tau_{UV}} d\tau\frac{ e^{-\tau M^2}}{\tau^{3/2}}\theta_2(0,e^{-4\pi^2\tau T^2}),
\end{eqnarray}
where the Jacobi function is defined as $\theta_2(0,q)=2\sqrt[4]{q}\sum\nolimits_{n=0}^\infty q^{n(n+1)}.$
Then the constituent quark mass  is
\begin{eqnarray}
M=m+\frac{6GMT}{\pi^{3/2}}\int^  \infty_{\tau_{UV}} d\tau\frac{ e^{-\tau M^2}}{\tau^{3/2}}\theta_2(0,e^{-4\pi^2\tau T^2}).
\end{eqnarray}

At finite volume, the quark momentum is  discretized and the integral over all spatial momenta is replaced by a sum over discrete momentum modes.
The discrete momenta  depending on the boundary conditions are
\begin{eqnarray}
 \vec{p}_{PBC}^2&=&\frac{4\pi^2}{L^2}\sum\nolimits_{i=1}^{3}n_i^2, \qquad n_i=0,\pm 1,\pm 2...\\
\vec{ p}_{APBC}^2&=&\frac{4\pi^2}{L^2}\sum\nolimits_{i=1}^{3}(n_i+\frac{1}{2})^2,\quad n_i= \pm 1,\pm 2... \label{eq.papbc}\\
 \vec{ p}_{SWC}^2&=&\frac{\pi^2}{L^2}\sum\nolimits_{i=1}^{3}n_i^2,\qquad n_i=+1,+2,+3...  \label{eq.psws}
 \end{eqnarray}
 where $L$ is the cubic volume size. The integration measure is replaced by sum of  discrete momenta
\begin{equation}
\int  dp (\cdots)\rightarrow\frac{2\pi}{L}\sum_{ n_i  }(  \cdots).
\end{equation}
Then the constituent quark mass is constrained by
\begin{eqnarray}\label{eq.gap3}
M&=&m+48GM\int^ \infty_{\tau_{UV}}  d\tau e^{-\tau M^2}[T \times  \nonumber \\
& &\sum^\infty_{k=-\infty}e^{-\tau \omega_k^2}\prod \limits_{i=1}^3 \sum_{n_i  }  e^{-\tau   p_i^2}]  \nonumber  \\
&=&m+48GMT\int^ \infty_{\tau_{UV}} d\tau e^{-\tau M^2}\times  \nonumber \\
&&\theta_2(0,e^{-4\pi^2\tau T^2}) [\frac{f(\theta)}{L}]^3,
\end{eqnarray}
with
\begin{eqnarray}
  { f(\theta)  }&=&\left\{ \begin{array}{l}
 {
   \theta_2(0,e^{-4\tau \pi^2/L^2})   \qquad \quad \quad \text{for APBC;}}  \\
   {\theta_3(0,e^{-4\tau \pi^2/L^2})  \qquad \quad \quad \text{for PBC;}} \\
    {\theta_3(0,e^{-4\tau \pi^2/L^2})-1  \  \quad \quad \text{for PBC-0;}} \\
 {[\theta_3(0,e^{-\tau \pi^2/L^2})-1]/2 \ \quad \text{for SWC.}}      \\
 \end{array} \right.
\label{psix}
\end{eqnarray}
Here we use PBC-0 to represent PBC without the zero-mode contribution (PBC would require an additional explicit treatment of the fermionic zero mode) and the $\theta_3(0,q)=1+2\sum\nolimits_{n=1}^\infty q^{n^2}$.

 \begin{figure}
   {\includegraphics[width=0.48\columnwidth]{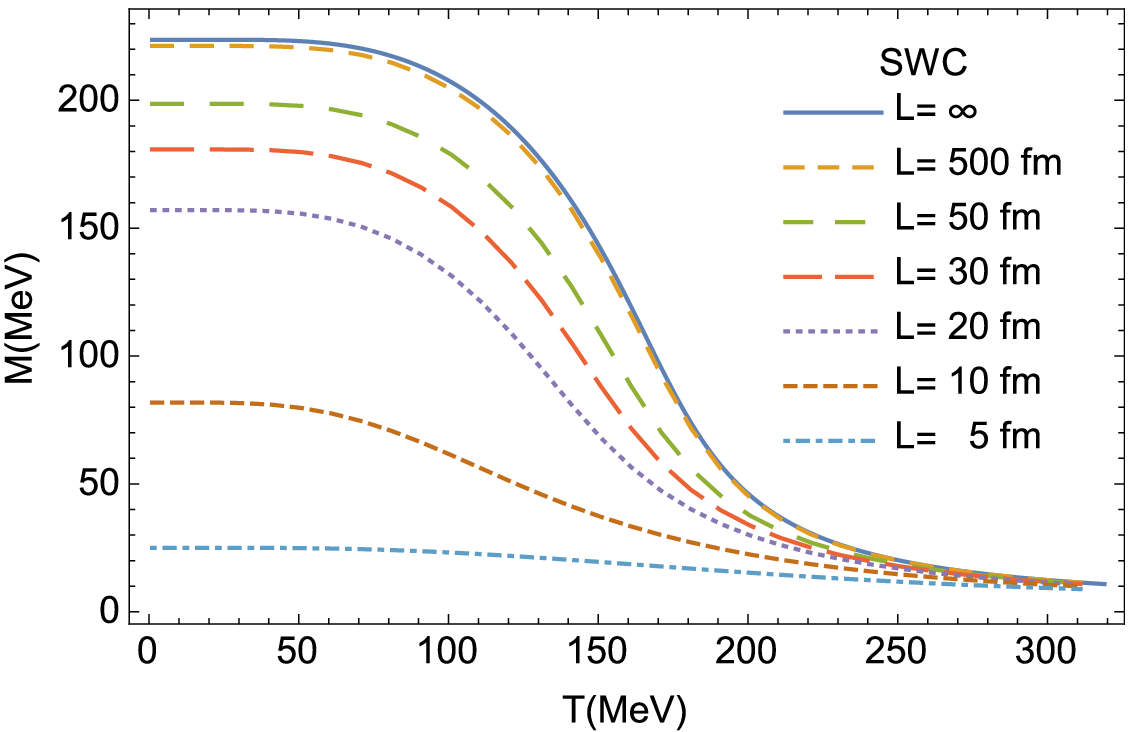} }
   {\includegraphics[width=0.48\columnwidth]{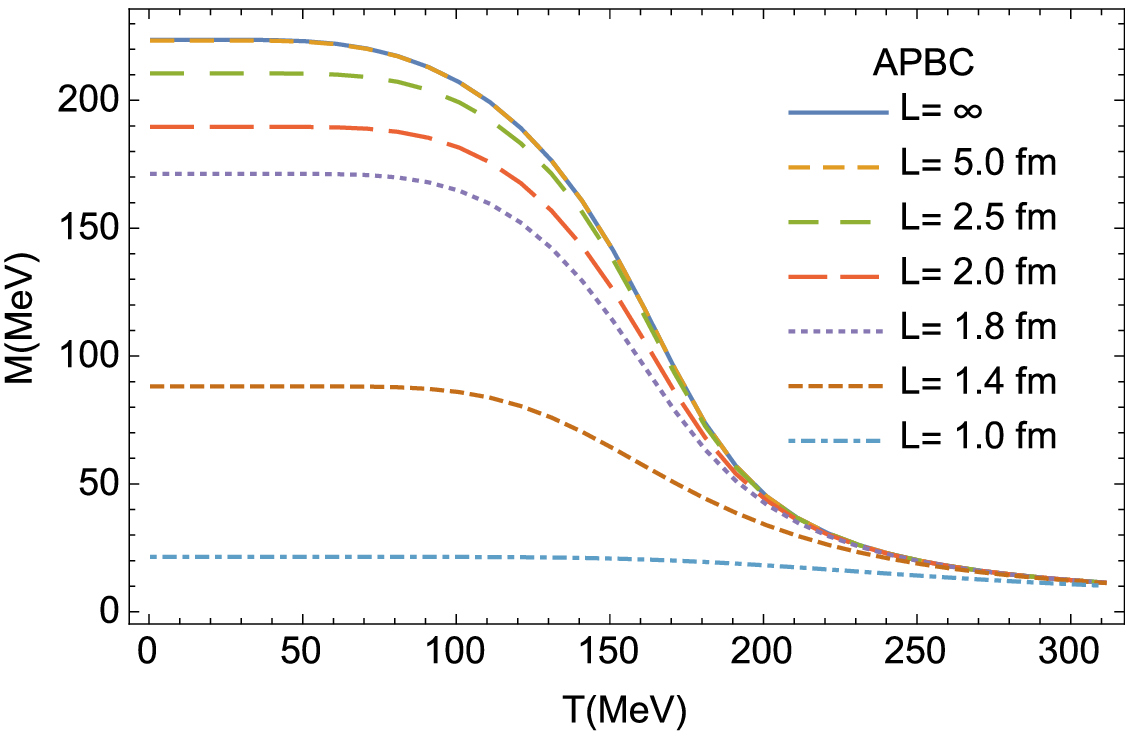} }  \\
  {\includegraphics[width=0.48\columnwidth]{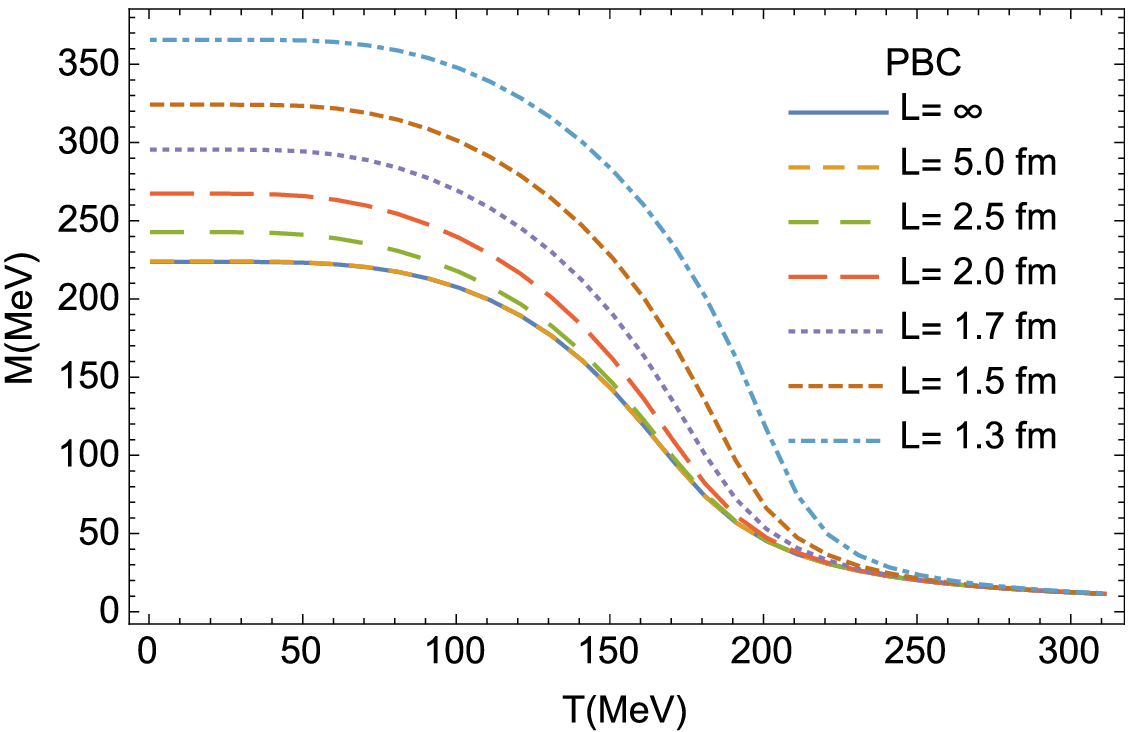} }
  {\includegraphics[width=0.48\columnwidth]{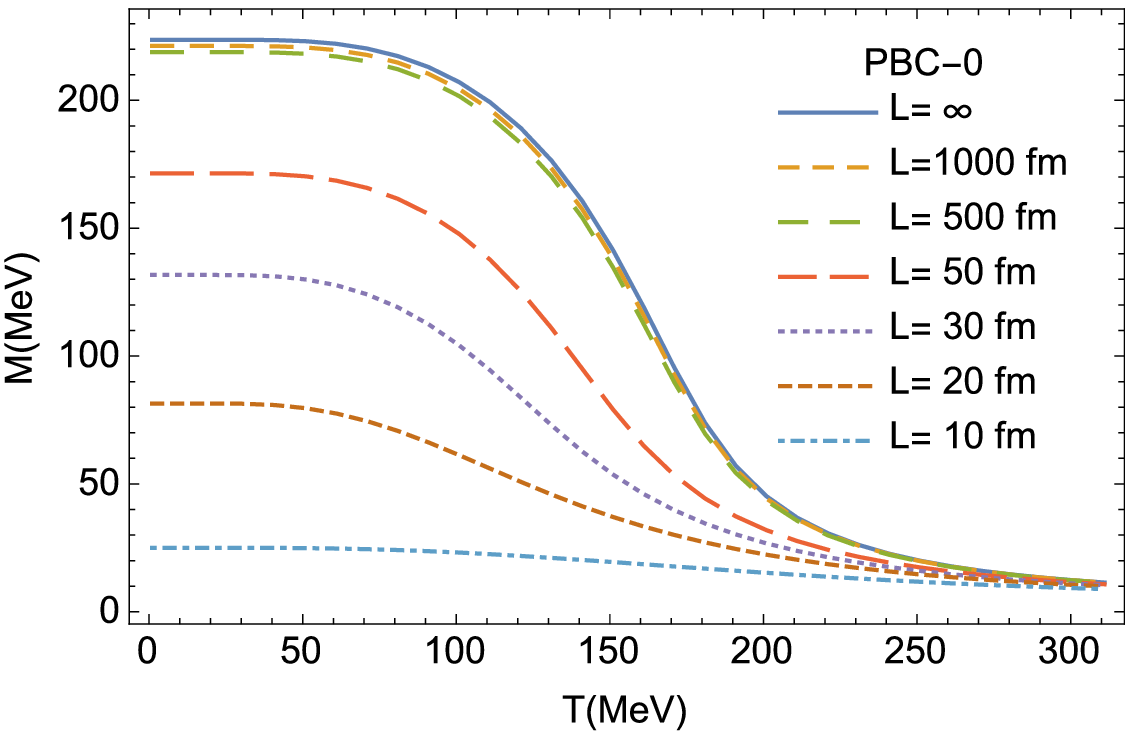} }\\
\caption{ Quark mass as a function of temperature and $L$.
 In contrast to SWC and APBC, the quark mass obtained from PBC increases as the volume decreases, owing to the zero-momentum contribution.}\label{fig.mass}
\end{figure}

The parameters we used here are $m=5$ MeV, $G=3.26*10^{-6}$ MeV$^{-2}$, $ \Lambda_{UV}=1080$ MeV and the $\tau_{UV}$ is given by $1/\Lambda_{UV}^2$.  With these parameters the quark mass is $ M=223.7$ MeV at zero temperature. In all the  calculations we neglect the possible dependence of the coupling on temperature and condensate which is discussed in Refs. \cite{Cui:2013aba,Wang:2016fzr, Ayala:2016bbi}. Also the effective coupling constants does not depend on the volume size \cite{Gasser:1987ah} in this work.

The quark mass $M$ at different $L$ and temperature under different boundary conditions  are plotted in Fig.(\ref{fig.mass}), with a few things noticeable. Firstly, for SWC, APBC and PBC-0,  the figures show crossover of chiral phase transitions. When the volume size $L$ is not very small, the quark mass or the chiral quark condensate smoothly reduces as the temperature increases.
 The three boundary conditions give same results in the thermodynamic limit. In addition, for APBC, when $L\geq L_0=5 $fm ($L_0$ is called the limit size)), the quark mass is indistinguishable from that at $L=\infty$ which is consistent with result from Ref.\cite{Luecker:2009bs}. While for SWC, it is found that the corresponding  $L_0 = 500$ fm. Both APBC and SWC the quark mass  deceases  as volume size decreases. These results are qualitatively consistent with those from Dyson-Schwinger equation with APBC \cite{Luecker:2009bs,Li:2017zny}. Secondly, for PBC, when the volume size $L$ decreases, the quark mass increases which is totally different with results from the other two boundary conditions. This is an effect of fermionic zero mode that is present for PBC.						
Thirdly, when the volume size $L\leq 2 fm$, there is only Wigner-Weyl solution for the case of SWC, where the  dynamic chiral symmetry is total restored. But, for PBC, The dynamical chiral symmetry breaking always exists for arbitrary $L$ at zero temperature. It is obvious that the choice of boundary conditions has a significant effect on the finite-volume mass shift.

The zero momentum contribution to the quark mass with PBC can be observed  through the gap equation Eqs.(\ref{eq.gap3}) and (\ref{psix}). The term $f(\theta)/L$ diverges as $L  \rightarrow 0 $ only for   PBC with a zero mode.  Without mechanism to restore the chiral symmetry, quark mass $M$ is nonzero and chiral condensate from the gap equation has a solution of negative infinity at low temperature.    At very high temperature, the zero momentum contribution is  heavily suppressed by the term $\theta_2(0,e^{-4\pi^2\tau T^2})$ and then the dynamical chiral symmetry gets restored.

\begin{figure}[h]
   {\includegraphics[width=0.7\columnwidth]{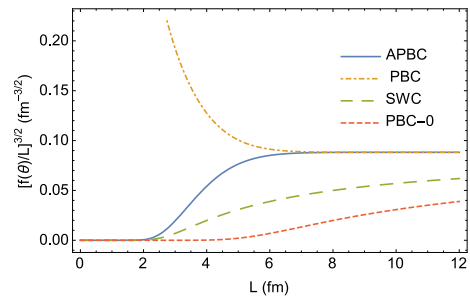} }
\caption{The  momentum summation as a function of volume size $L$ for different boundary conditions.}\label{fig.theta}
\end{figure}

 The reason of these differences for  different choice of boundary condition can be illustrated by Fig.(\ref{fig.theta}). It is a plot of function $[f(\theta)/L]^{3/2}$ with $k=\tau \pi^2=20$. We have found that for any value of $k$, the relative positions of those curves keep invariant.  For a fixed $L$,  the order of  curves along the vertical axis of  Fig.(\ref{fig.theta}) is also the order for quark mass in different boundary condition. From Eq.(\ref{eq.gap3}), as M is much large than current quark mass $m$ and temperature is fixed, we reach to a relation
 \begin{equation}
 e^{\sqrt{\tau}M}\sim[\frac{f(\theta)}{L}]^{\frac{3}{2}}.
 \end{equation}

 In  Fig.(\ref{fig.theta}), only the curve of PBC is a monotonically decreasing function of $L$. This explains why the effective quark mass increases  as  $L$ decreases for PBC.  The curves of PBC and APBC always reach the same limit as $L$ increases and to approach their thermodynamic limit. The $L_0$ for PBC or APBC is much smaller than the one for  SWC and PBC-0, which can explains the different behaviors of quark mass with different boundary conditions.  The large value of  function $[f(\theta)/L]^{3/2}$  means more quark field momenta  are ``squeezed" in unit volume which leads to an increase of chiral condensate and then  constituent quark mass.

 The crossover behavior of quark chiral condensate can be depicted by the
  chiral quark condensation with respect to temperature and current quark mass. The chiral quark condensation with respect to temperature is  defined as
\begin{eqnarray}\label{ }
 \chi_T(T)&=&-\frac{\partial \sigma}{\partial T}.
\end{eqnarray}

\begin{figure}
{\includegraphics[width=0.48\columnwidth]{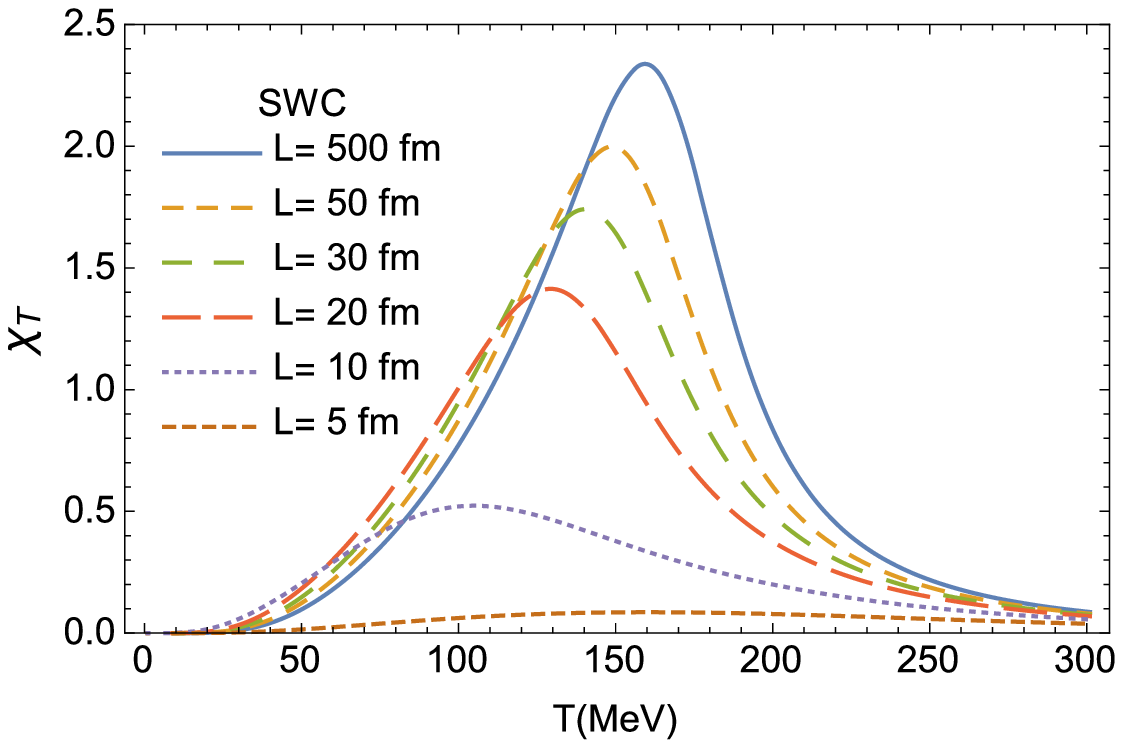} }
{\includegraphics[width=0.48\columnwidth]{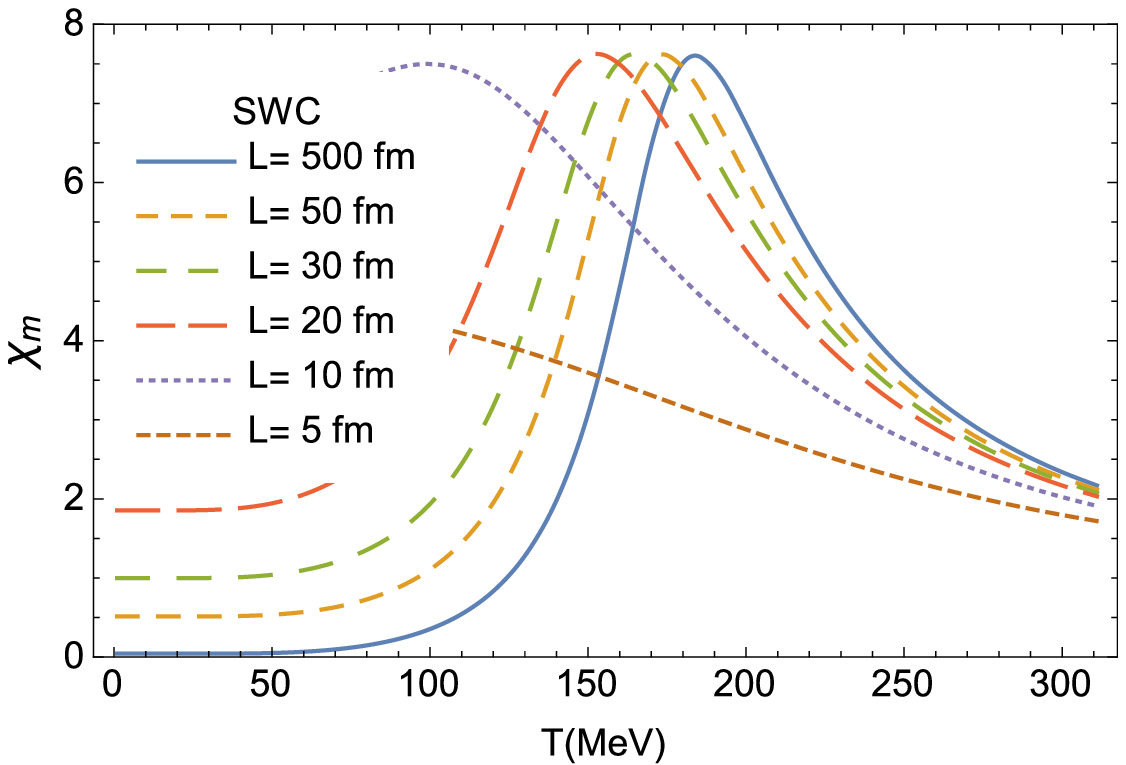} }\\
{\includegraphics[width=0.48\columnwidth]{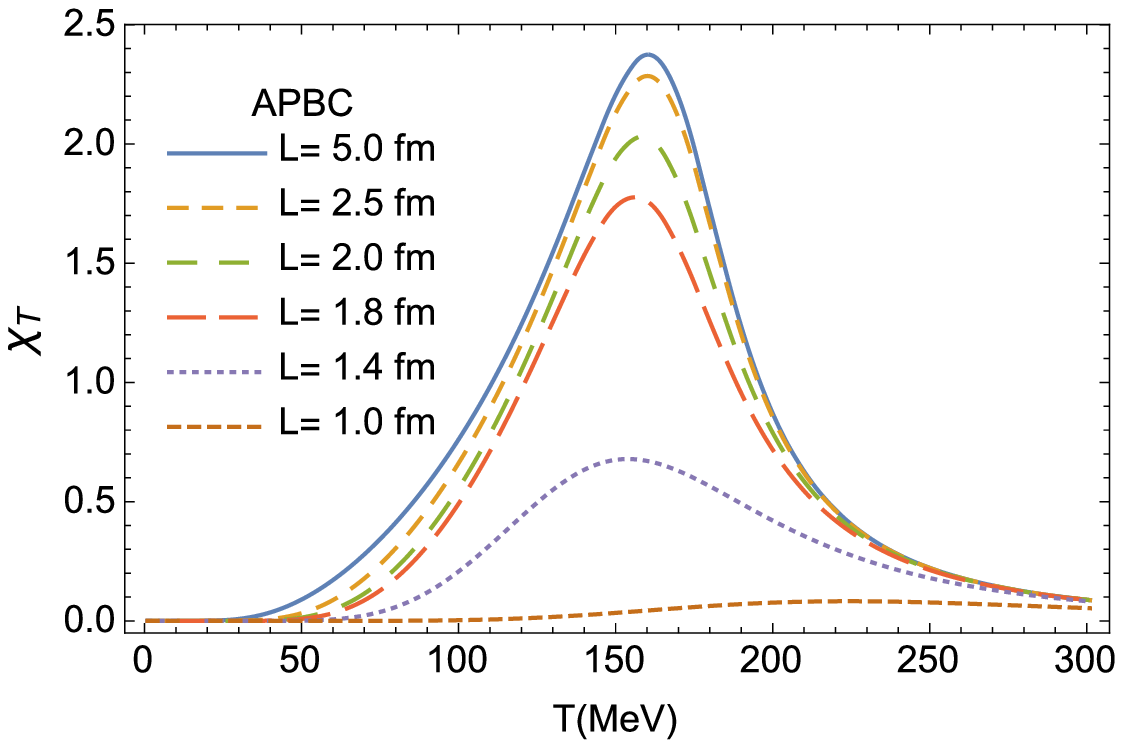} }
{\includegraphics[width=0.48\columnwidth]{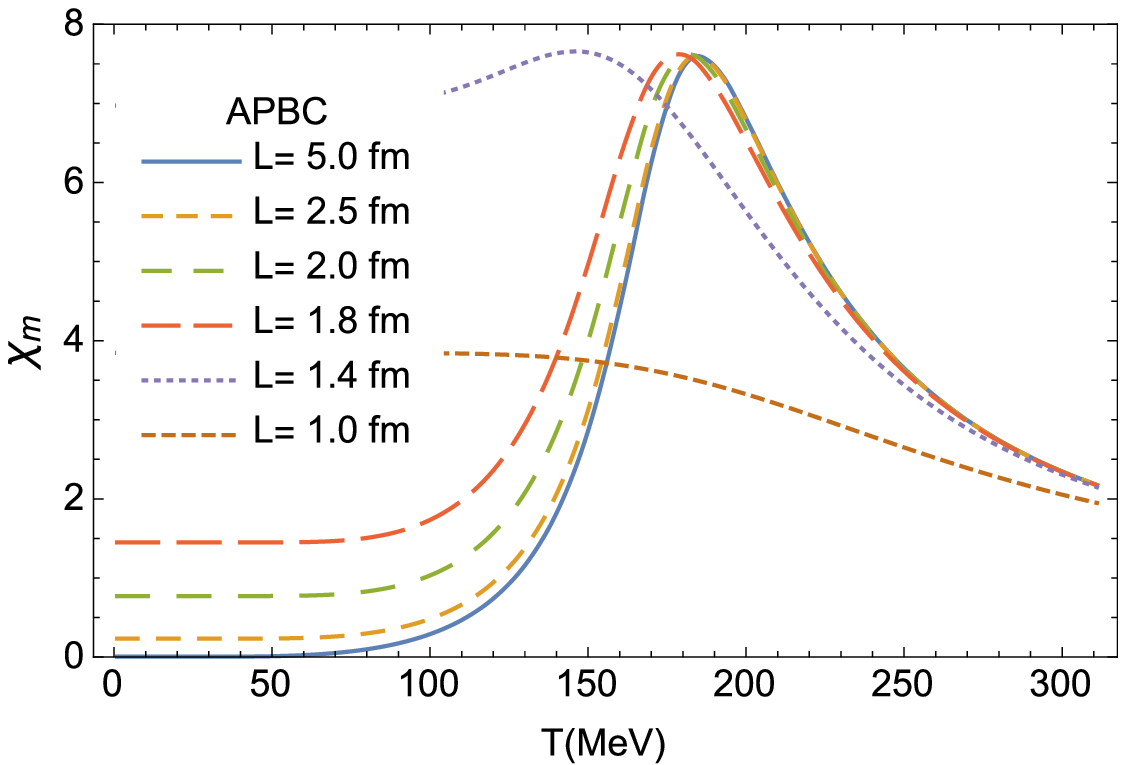} }\\
{\includegraphics[width=0.48\columnwidth]{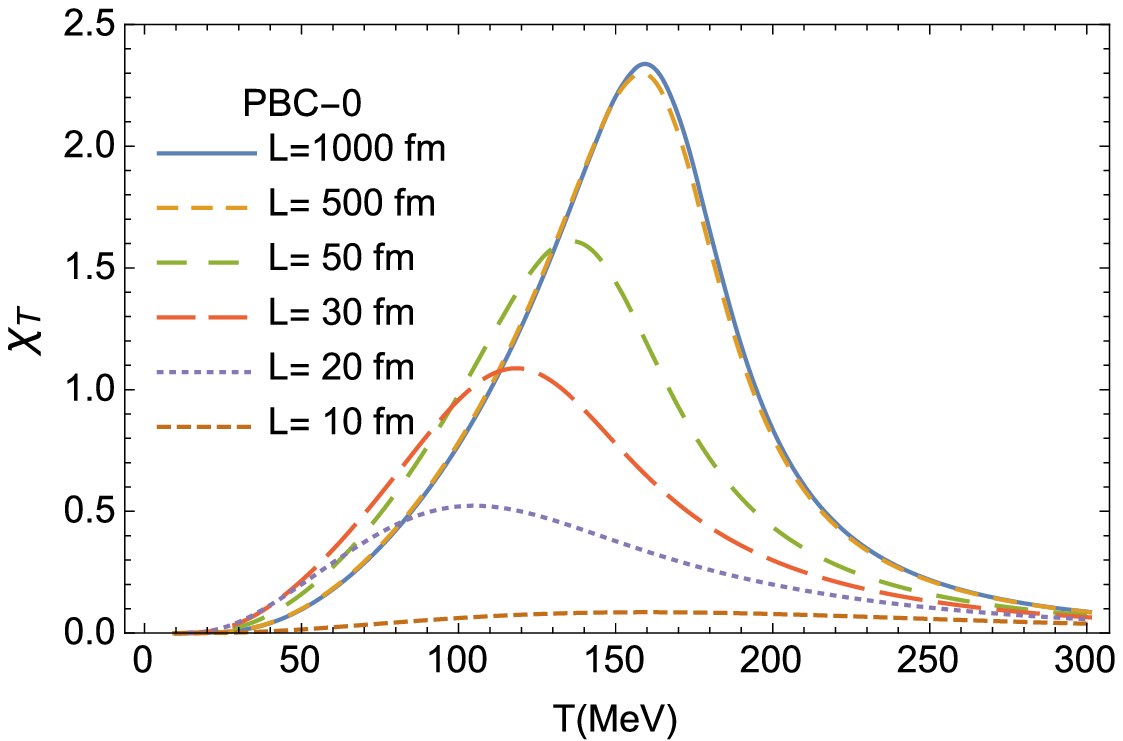} }
{\includegraphics[width=0.48\columnwidth]{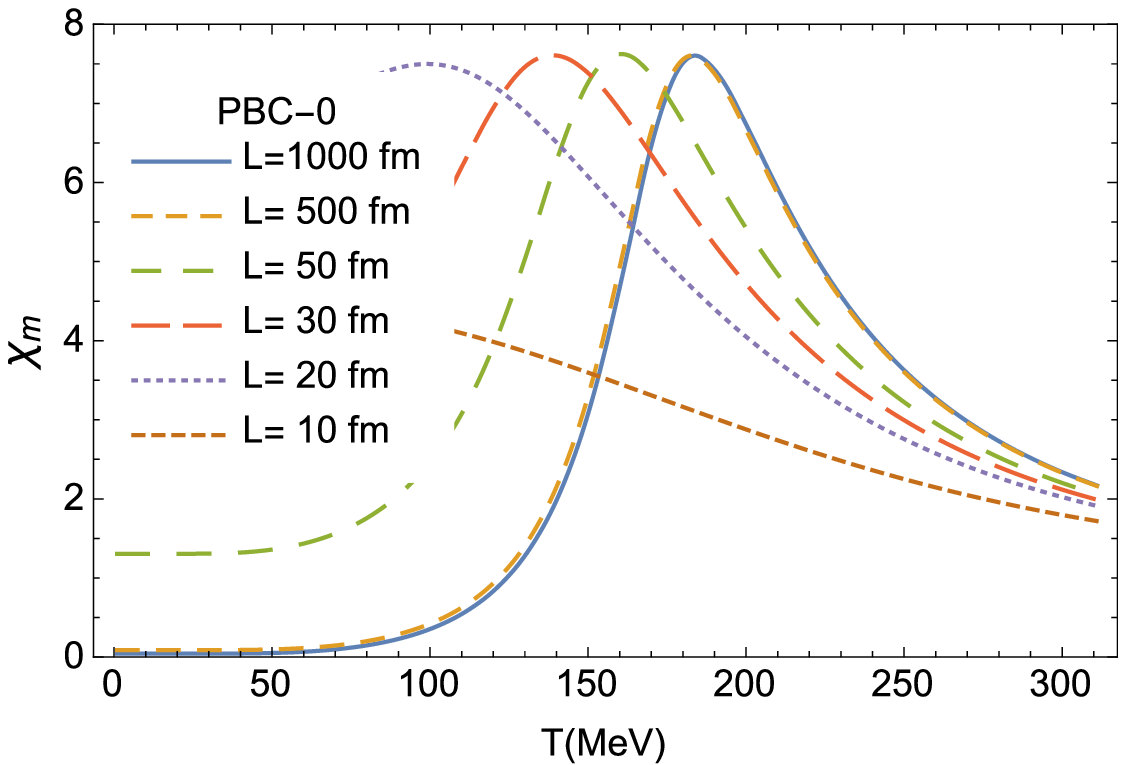} }\\
\ {\includegraphics[width=0.48\columnwidth]{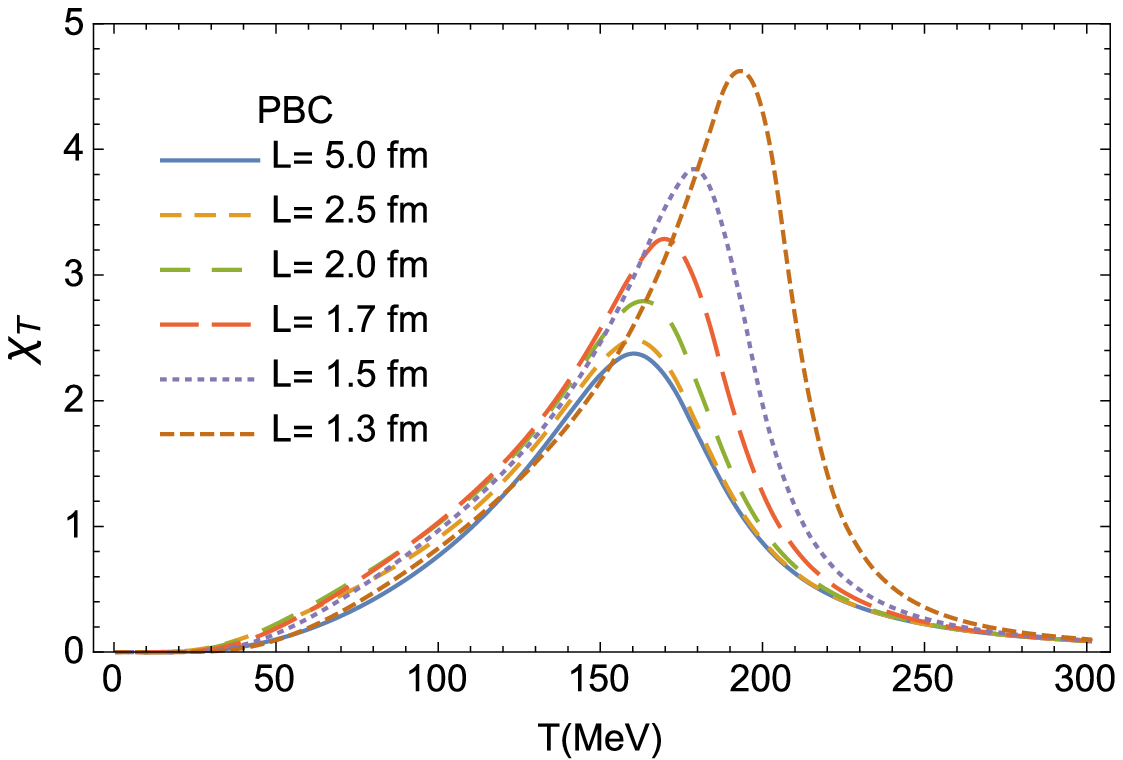} }
{\includegraphics[width=0.48\columnwidth]{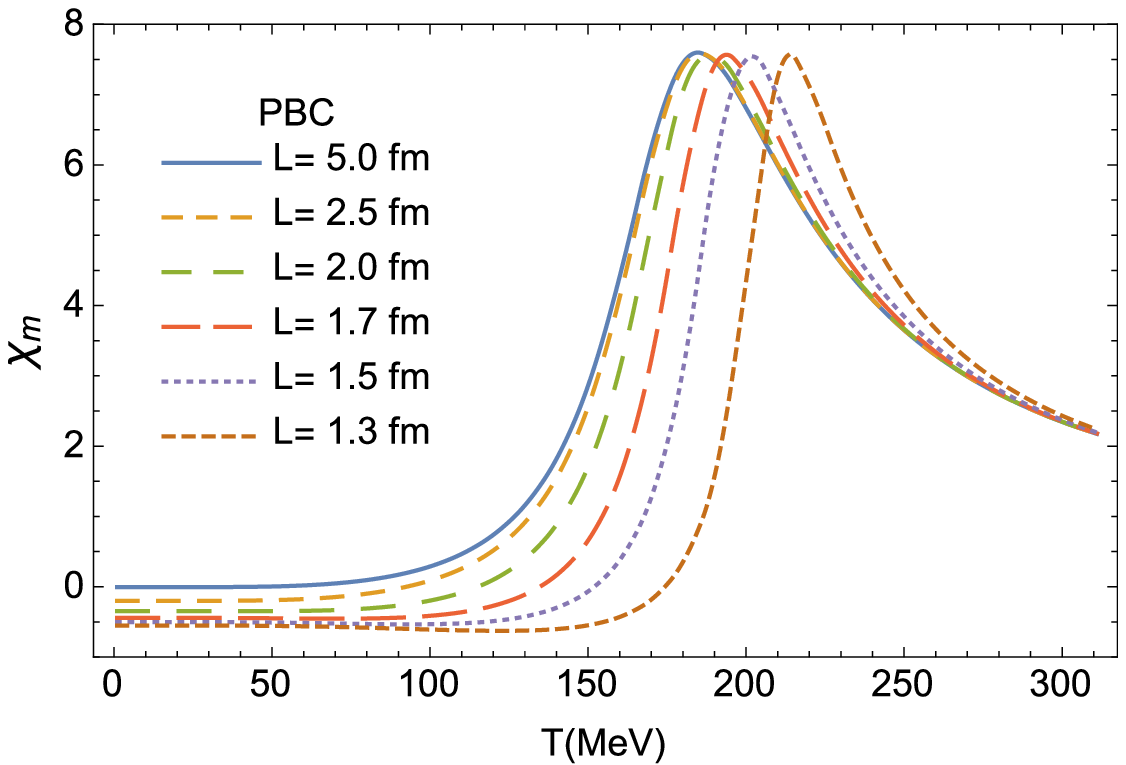} }\\

\caption{ Chiral susceptibilities as function of temperature at different volumes. $\chi_T$  and $\chi_m$ are the  susceptibilities with respect to temperature and current quark mass respectively. }\label{fig.xt}
\end{figure}

The susceptibility with respect to current quark mass can be easily  derived from Eq.(\ref{eq.gap3}) as all the parameters are fixed.  But direct derivation of  Eq.(\ref{eq.gap3}) give  $\chi_m(T)$ nonzero in the infinite volume at zero temperature.  Actually, in the proper time regularization, parameters $m$, $G$ and $\Lambda_{UV}$ are fixed by experimental values of decay constant and mass of pion. Therefore if the coupling $G$ is fixed, the ultraviolet cutoff  $\Lambda_{UV}$ must have dependence on the current quark mass $m$. In this consideration, we give a small change $\delta_m$ to $m$ and get new cutoff $\Lambda_{UV}(m+\delta_m)$. Then  the $\chi _m(T)$  can be  deduced from formula
\begin{equation}
 \chi_m(T)=\frac{\sigma(m+\delta_m)-\sigma(m)}{\delta_m}.
 \end{equation}

The results for the susceptibilities are showed in Fig.(\ref{fig.xt}).  The two kinds of susceptibility have different behaviors  when the volume size $L$ becomes very small.  At zero temperature,  the susceptibility $\chi_T$ is zero in any boundary condition and unaffected by the size of the boundary, but the susceptibility $\chi_m$  increases (decreases) as $L$ decreases for APBC (PBC). Note that   $\chi_m$ changes the sign as $L$ decreases for PBC.

In the infinite volume limit and chiral limit, the sharp peak in the  chiral susceptibility plot define the phase transition point. While beyond the chiral limit, we still take the  chiral susceptibility as the order parameter and use the maximum to find the pseudo-transition temperature.
The pseudo-critical temperatures defined from the susceptibilities are   denoted  as $T_c^m$ and $T_c^T$.
 In the infinite volume limit, $T_c^m\simeq 165 $MeV  and $T_c^T\simeq 185$ MeV.  This difference also exists in finite volume. Tab. {\ref{tab.tc}} shows the pseudo-critical temperature at different volumes. For a specific boundary condition, the  pseudo-critical temperatures $\chi_m$ and $\chi_T$ respond to the volume size in the same way.
  For SWC, the  $T_c^m$ and $T_c^T$ decrease with smaller $L$  while for PBC the  $T_c^m$ and $T_c^T$ increase as $L$ decreases. For APBC, $T_c^m$ and $T_c^T$ only slightly decreases with the decrease of $L$ which is  consistent with the result in Ref. \cite{Braun:2005fj} with large current quark mass.

\begin{table}
  \caption{Pseudo-critical temperature deduced from the susceptibilities $\chi_m(T)$ and $\chi_T(T)$. The temperatures are in unit MeV and volume size $L$ is in unit fm. }
 \label{tab.tc}
 \begin{center}
 \begin{tabular}{c c c c c c c}
    \hline
 SWC& $L$ &500&50&30&20&  \\
  &$T_c^m$ & 184 & 173  &165&153\\
  & $T_c^T$&  164 &  153& 145&134\\
   \hline

   PBC-0& $L$ &500&50&30&20&  \\
 & $T_c^m$&  183 &  160& 139&93\\
  &$T_c^T$ & 163 & 142  &118&94\\
     \hline
 APBC&$L$&5.0&2.5&2.0&1.8\\
 & $T_c^m$ & 185 &  184 & 182 & 179\\
 &  $T_c^T$&  165 &  165&  163 & 161 \\
   \hline

 PBC &$L$&5.0&2.5&2.0&1.7\\
 & $T_c^m$ &  185& 186&  188&  194 \\
 &   $T_c^T$&  165 & 166 &  168 & 175 \\
    \hline
 \end{tabular}
\end{center}

\end{table}

\begin{figure}[h]
{\includegraphics[width=0.48\columnwidth]{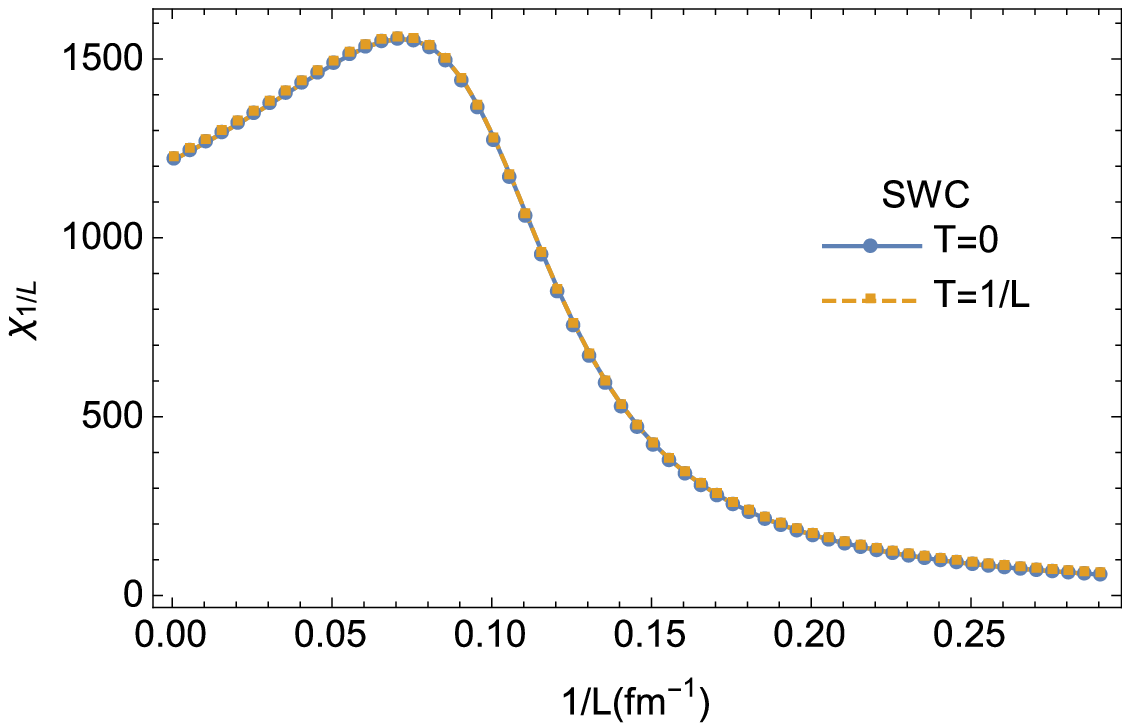} }
{\includegraphics[width=0.48\columnwidth]{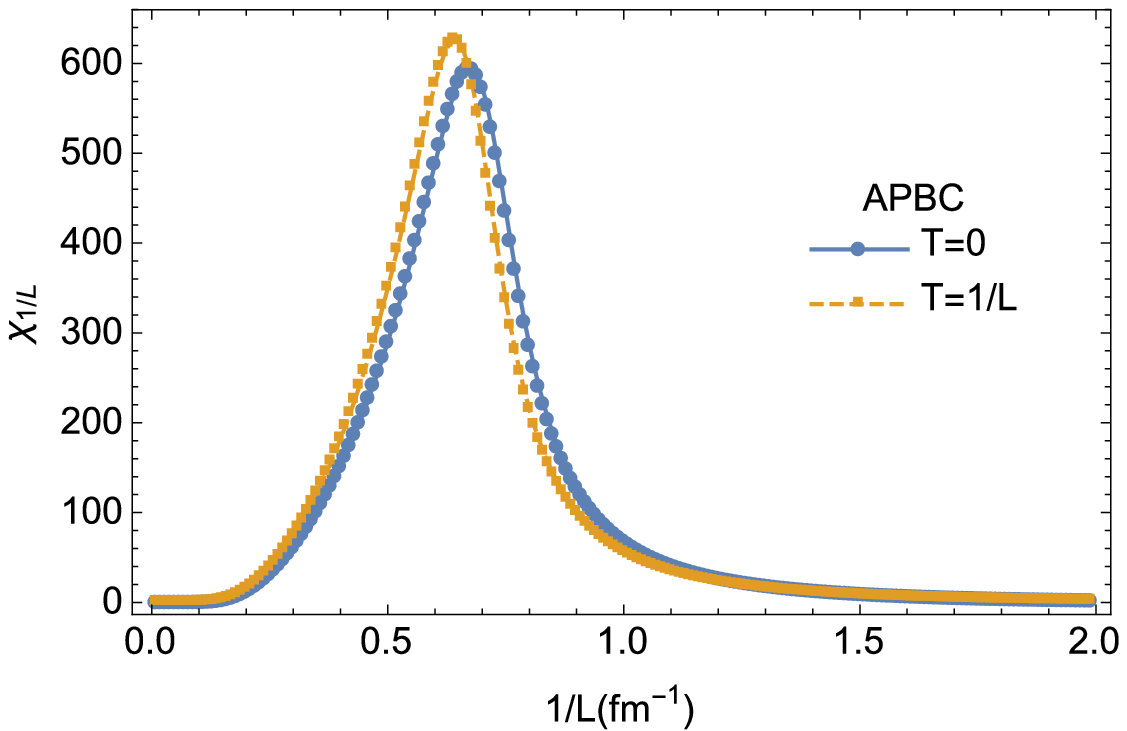} }\\

\caption{ The  vacuum susceptibility with respect to volume size $1/L$.  For quark field for SWC, the $T=0$ and $T=1/L$ curves coincide and the dynamical chiral symmetry is restored at $L=2$ fm.     }\label{fig.Lxas}
\end{figure}
In order to quantitatively reflect the finite volume effects on the QCD chiral phase transition, similar to the  chiral quark condensation with respect to temperature, here we introduce a new vacuum susceptibility,  which is defined as the derivative of the  chiral quark condensation with respect to spatial size $1/L$. We call it spatial susceptibility which reads as
\begin{equation}
\chi_{1/L}(T)=-\frac{\partial \sigma}{\partial (1/L)}.
\end{equation}

According to the illustration, see Fig. (\ref{fig.Lxas}), it is emphasized that in the Euclidean space, the discretization in the temporal direction (temperature $T$) and the discretization in the spatial direction $(1/L)$ are equivalent. That is, as the temperature $T$   or $1/L$ increases, chiral symmetry  will be partially restored.

In summary, we have used the NJL model to study the chiral crossover transition in a finite volume. Besides the two commonly used APBC and PBC, we have chosen the SWC for the quark field as a real physical boundary condition. It is found that different boundary choice for the finite volume  has dramatically influence on the QCD  chiral behavior. Starting from the infinite volume,  only PBC gives quark mass that increases as  the volume size decreases and the chiral susceptibility $\chi_m(T)$ become negative at low temperature.  The strange behavior of chiral quark condensate for PBC is due to dominant contribution from the zero mode at small $L$. In order to avoid this strange behavior, we use PBC-0 to represent the  period boundary condition without the zero-mode contribution. Finally, we found that the results from PBC-0 are similar to that from SWC.

Here it should be noted that the results by means of SWC are obviously different to the results from the APBC and PBC.  Although the three boundary conditions give the same chiral crossover transition curve in the infinite volume limit, the limit size $L_{0}$ using SWC is  $L_{0}\approx 500$ fm which is much larger than the results obtained using PBC or APBC. Especially important,  $L_{0}\approx 500$ fm is also far greater than the current maximum size in lattice simulations of full QCD in numerical calculations.

In the past it is hard to conceive of systems that are small enough to lead to observable finite-volume effects, since the length scales involved are so small compared to the typical extent of the system.
However the experiment with relativistic heavy ion collisions has changed a lot. At present the estimated volume of  QGP in RHICs is $\sim 250$ fm $^3$ which can be compared with the wave length of the $\pi$ meson. This means that the finite volume effects should be observed experimentally in RHICs. According to our calculation, the finite-volume effects may play a significant role in the  QGP. Furthermore, a new spatial susceptibility reflecting  the finite-volume effects of the chiral restoration in QGP was introduced and it was found that the finite-volume effects and the temperature effects were completely equivalent, namely for very small volume size or large temperature, chiral symmetry is effectively restored.

\bigskip
This work is supported in part by the National Natural Science Foundation of China (under Grants No.11475085,  No.11535005 and 11690030).

\end{document}